# A Piggybank Protocol for Quantum Cryptography

Navya Chodisetti

**Abstract**
This paper presents a quantum mechanical version of the piggy-bank cryptography protocol. The basic piggybank cryptography idea is to use two communications: one with the encrypted message, and the other regarding the encryption transformation which the receiver must decipher first. In the quantum mechanical version of the protocol, the encrypting unitary transformation information is sent separately but just deciphering it is not enough to break the system. The proposed quantum protocol consists of two stages.

**Introduction**
There exist at least two separate paradigm of the secure communication system from the perspective of the usage of keys: (i) the standard cryptographic scheme where the key is required at both ends (and the key can be symmetric or non-symmetric) [1], and (ii) where the key information is sent separately using a different communication in such a manner that its decipherment by the eavesdropper in itself does not break the communication (piggybank protocol of [2]). The piggybank paradigm breaks up the secret communication into two parts somewhat in the spirit of recursive hiding of secrets [3], the interlock protocol [4] and the ElGamal cipher [5], excepting that it considers the transformation in a more general setting. This potentially makes the task of the eavesdropper harder because she must break both the messages correctly in order to break the system. It also makes it possible to trade off the complexity of the two transformations in a manner that makes it suitable for use in some compute-capacity limited applications.

Here we present a simplified implementation of the three-stage protocol [6] using the piggy-bank idea. This protocol has been implemented [7],[8]. The motivation is to reduce the number of stages of the protocol. The piggybank version of the protocol reduces to two stages which can be of value in several applications.

The paper firstly describes the piggybank idea in a classical setting by means of an example. Then the quantum cryptographic protocol is presented.

**Background: The basic Piggy bank protocol**
The basic classical piggybank protocol [2] introduces an additional element in the communication process. Bob sends an empty locked piggy bank to Alice. When she receives it,

Alice deposits the secret into the box together with the decryption key of a coded letter. In addition, she prepares a letter to be sent separately. The piggy bank and the letter are sent back to Bob.

The letter is required to authenticate the contents of the locked piggybank box. It cannot be in plain text because the content list itself is a secret. Bob opens the box, obtains the key and then unlocks the secret within the letter using the key. In this implementation of the piggybank protocol for data, Bob obtains secret message (S) and key which is h(S).

**Step1**. Bob starts with a random number R and the piggy bank transformation is represents by a one way transmission f(R)= $R^e$ mod n, where n is a composite number with factors known only to him; e is the publicly known encryption exponent.

**Step2**. Bob sends f(R) to Alice who multiplies it with hash function h(S) which is denoted as K. Alice sends $K(R^e) + S \mod n$ to Bob in one communication and f(K)= $K^e \mod n$ in another communication.

**Step3**. Bob uses his secret inverse transformation to first recover K and having found it he can recover S.

**Example 1**: Let n=522617 and the public encrypting exponent is e=5 (with the secret decrypting algorithm being 416,861). Bob chooses random R=1201 and sends $1201^5$ mod 522617= 169841 to Alice.

Alice's random secrets are S=11925 hash function of S is h(S) i.e. K=5. Alice computes $169841 \times 5 + 11925 = 861130$ and sends it and also $5^5$ mod 522617 = 3125 to Bob.

Bob uses his secret decryption exponent to recover K: $3125^{416861}$ mod 522617 = 5. Thus $5 \times 169841 + S = 861130$, from which he recovers S.

**Example 2**. Let n=124711 and the public encrypting exponent is e=3 (with the secret decrypting algorithm being 82667). Bob chooses random R=2101 and sends $2101^3$ mod1 24711 = 102786 to Alice.

Alice's random secrets are S=9278 hash function of S is h(S) i.e. K=8 Alice computes

$102786 \times 8 + 9278 = 831566$ and sends it and also $8^3$ mod 124711 = 512 to Bob.

Bob uses his secret decryption exponent to recover K: $512^{82667}$ mod1 24711 = 8. Thus $8 \times 102786 + S = 831566$, from which he recovers S.

Assuming that this protocol uses a cryptographically strong random number generation algorithm, it makes it difficult for the eavesdropper to guess the value of R and break the public encryption key e. Apart from random number generation use of a complex hash function will also prevent the protocol from Man in the middle attack. Even if Eve tries to attack and gets hold of the secret message, Alice and Bob can introduce classical authentication resource say one-time pad, which might be able to recognize the misinformation from Eve, if any.

## The Proposed Quantum Protocol

Consider the arrangement of Figure 1 to transfer state Y from Alice to Bob. To transfer state Y from Alice to Bob the state Y is one of the two orthogonal states, such as $|0\rangle$ and $|1\rangle$, or $\frac{1}{\sqrt{2}}(|0\rangle+|1\rangle)$ and $\frac{1}{\sqrt{2}}(|0\rangle-|1\rangle)$, or $\alpha|0\rangle+\beta|1\rangle$ and $\alpha|0\rangle-\beta|1\rangle$ The orthogonal states of Y represent 0 and 1 by prior mutual agreement of the parties.

Alice and Bob apply secret transformations $U_A$ and $U_B$ which are commutative, i.e $U_A U_B = U_B U_A$. An example for this would be $U_A = R(\theta)$ and $U_B = R(\phi)$, where $R(\theta) = \begin{bmatrix} \cos\theta & -\sin\theta \\ \sin\theta & \cos\theta \end{bmatrix}$. In other words, rotations are limited to linear polarization. The sequence of operations in the protocol is as follows:

**Step 1**.
Bob applies the transformation $U_B$ on X, a random polarization state which is the *cover* state, and sends *n* qubits of it to Alice.

**Step 2**.

2.1 Alice applies $U_A$ on the received *n* qubits to form $U_A U_B(X)$ and sends them bank to Bob in one communication.

2.2 Somewhat later she sends $U_A^+(Y)$ in another communication in *m* qubits where Y is the secret message bit ($m << n$).

**Step 3**.
Bob applies $U_B^+$ on the received bunch of *n* cover qubits and then performs tomography to get the transformation $U_A$ and then he applies this transformation on the second received communication to get secret message bit Y.

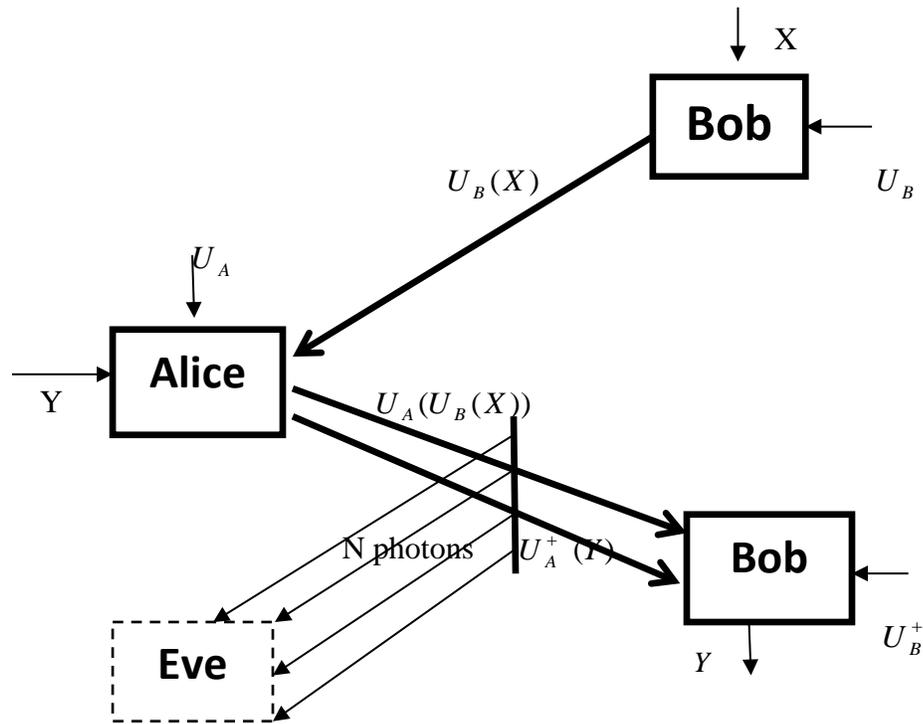

Figure 1. Piggybank protocol for quantum cryptography

The transformations used by both Alice and Bob to send their qubits are secret (although Alice's rotation is deciphered by Bob before he can obtain the value of the message bit Y). To retrieve secret message bit Y from $U_A^+(Y)$ it is necessary for Bob to calculate $U_A$. It is possible only if they transmit multiple photons and calculate $U_A$ and then apply it to the inverse message to get Y. Here X is random information that remains known only to Bob.

This protocol assumes that the eavesdropper cannot estimate the original polarizations [9]-[11] and, of course, due to the no-cloning theorem he cannot make multiple copies of the transmitted photons. If m =1, the system is provably secure.

To consider the value of n, let the tomographic scheme used by Bob extract $\log_2 n$ bits of information regarding the polarization angle or the transformation $U_A$. If Bob is going to choose out of k discrete angles that are equally spaced then $n = 2^k$.

The eavesdropper will need n photons to determine the angle. Therefore, the system would be safe as long as the number of photons being used is less than 2n. If Even siphons off too many photons, she would be detected.

Eve would need more then n photons to determine $U_A$. Let us consider that Eve needs $kn$ photons (where k>1) and Bob can send up to $kn$ photons.

There is also an inverse relationship between n and m. If a large n is used, m should be small and vice versa.

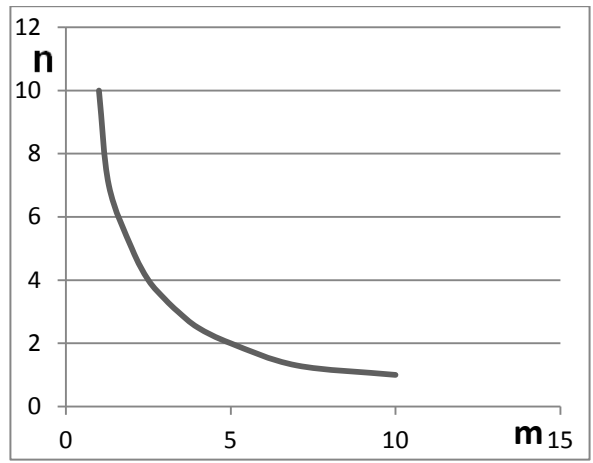

Figure 2: Inverse function between n and m for same error problem

Table 1 discusses the Game theoretic view of the problem where Eve is involved in the system. The system is meant to be safe until the number of photons is less than 2n. The challenge will be to use the inverse relationship between n and m to force Eve into siphoning few photons which will keep the system safe.

Table 1: Game theoretic view of the Problem

|  | Eve siphons few photons | Eve siphons many photons |
|---|---|---|
| Alice and Bob use low n, m | Safe but low range Number of photons <2n | Safe since Eve is Detected Number of photons <2n |
| Alice and Bob use high n, m | Safe Number of photons <2n | Unsafe Number of photons $\geq 2n$ |

## Conclusions

This note has presented a provably secure variant of the three-stage protocol where only two stages are required. The proposed system has two parameters, m and n, which are the number of photons being used for the cover qubit and the message qubit respectively. This provides flexibility as far as the usage of the system is concerned. We have also discussed the game theoretic approach of the system to Eve.